\documentclass[preprintnumbers]{revtex4}
\usepackage{graphicx}
\usepackage{dcolumn}
\usepackage{bm,hyperref}
\usepackage{amsmath}
\usepackage{amsfonts}
\usepackage{amssymb}
\usepackage[latin1]{inputenc}%
\begin{document}
\title{Suppression of light propagation in a medium made of randomly arranged
dielectric spheres}
\author{Carsten Rockstuhl\footnote{\textsf{carsten.rockstuhl@uni-jena.de}}}
\author{Falk Lederer}
\affiliation{Institute of Solid State Theory and Optics, Friedrich-Schiller-Universität,
Max-Wien-Platz 1, D-07743 Jena, Germany}

\begin{abstract}
Light propagation in a medium made of densely packed dielectric spheres is investigated
by using a rigorous diffraction theory. It is shown that a substantial suppression of the
local density of states occurs in spectral domains where the single constituents exhibit
Mie resonances. The local density of states decreases exponentially at the pertinent
frequencies with a linearly increasing spatial extension of the aggregated spheres. It is
shown that a self-sustaining random arrangement of core-shell spheres shows the same
fundamental characteristics. Such approach offers a path towards easy to fabricate
photonic materials with omnidirectional gaps that may find use in various applications.

\end{abstract}

\maketitle

In the past several years light propagation in random media composed
of discrete scattering objects has attracted a considerable deal of
interest. It derives its fascination not only from numerous
potential applications such as, e.g., the random laser
\cite{Wiersma_NatPhys,Wiersma_NatPhot}, but also because it
challenges our fundamental understanding of light propagation. Most
notably, the effect of light localization attracted considerable
attention \cite{McCallNature}. Light localization can be understood
as a synonym for the occurrence of a complete photonic band gap
where spectral components of light within the photonic band gap must
not propagate within the medium because Maxwell's equations provide
only evanescent wave solutions. By placing a source that emits light
at a frequency within the gap inside such material, the light
transport off the the source is suppressed. It remains localized.
Light localization in one-dimensional structures may be easily
achieved by, e.g., introducing a defect into a sequence of
dielectric layers arranged to form a Bragg stack. In two-dimensional
media made of infinitely extended cylinders of high permittivity
this localization was feasible by exploiting scattering resonances
of sufficient strength \cite{Zhang_2D,Rockstuhl_OL}. As a signature
of localization one may use the exponential decay of the local
density of states (LDOS) at a point inside the medium with
increasing size of the system. A significant suppression of the LDOS
in a finite system may be interpreted as a sign of the appearance of
a complete band gap, although it does not constitute a rigorous
proof. Because scattering resonances also affect the formation of a
gap for a periodic arrangement of the cylinders, a complementary
computation of the photonic band structure proved that the band gaps
occurs indeed in the same spectral domain as the reduced LDOS
\cite{Rockstuhl_OL}.

In three-dimensional systems, made of spheres \cite{Righini}, this
possibility of light localization is controversially discussed as in
experiments absorption may potentially shade localization effects
\cite{Absorption}. To investigate theoretically the possibility of a
suppression of light propagation, one may take advantage of an
analytical approach \cite{Rusek}. But as the structure involves a
high index contrast and a size of the scatterers that is comparable
to the wavelength, usually the problem requires a rigorous solution
of Maxwell's equations \cite{Sebbah,Ye}. Using a super cell approach
rigorous computations of the photonic density of states can be
performed. For an amorphous photonic material consisting of a
continuous-random-network a complete photonic band gap has been
identified recently \cite{Notomi}. This work has shown that
appropriate short-range order might also permit for light
localization, in addition to the well-known suppression of light
propagation in periodic media due to Bragg resonances \cite{John}.

However, most experimental investigations of light propagation in 3D random
media \cite{Lopez} employed sufficiently dense packed high permittivity
spheres. This is mainly due to their easy fabrication procedure that relies on
colloidal chemistry. The excitation of Mie resonances in these spheres is
usually regarded as the primary reason for suppressing propagation
\cite{Genack,Maret}. To theoretically verify the effects associated with light
localization, various efforts were undertaken in the past. In Ref.
\onlinecite{Cao} the density of states was computed for small clusters of
spheres using a generalized Mie theory. In Ref. \onlinecite{Mishchenko}
scattering effects depending on the cluster size were studied for low index
contrast using the superposition T-matrix method. A large scale
finite-difference time-domain (FDTD) method was also used to trace light
propagation in ensembles of spheres with a strongly varying size \cite{Conti}.

In this work we focus on investigating the LDOS of large clusters
made of randomly arranged identical high permittivity spheres. For
this purpose we use the FDTD method \cite{FDTD}. It is shown that
the LDOS decays exponentially with increasing size of the cluster in
certain frequency intervals indicating a significant reduction of
light transport velocity. The spectral positions of these domains
are correlated to Mie resonances. They occur independently of the
arrangement and are tunable by modifying the permittivity of the
spheres. However, below a certain contrast the frequency intervals
of the suppressed LDOS are hardly to recognize. We also investigate
the LDOS of core-shell spheres which were arranged by using the
sphere dropping method. Consequently they form a self-sustaining
ensemble of spheres. The approach allows to maintain the main
optical characteristics if the permittivity of the shell is
sufficiently low. This work opens a path towards an easy to
fabricate photonic material that allows for a suppression of light
propagation regardless of its direction and polarization. Such a
material can be applied as, e.g., isotropic filters for solar cells.

\begin{figure}[h]
\centering
\includegraphics[width=180mm,angle=0] {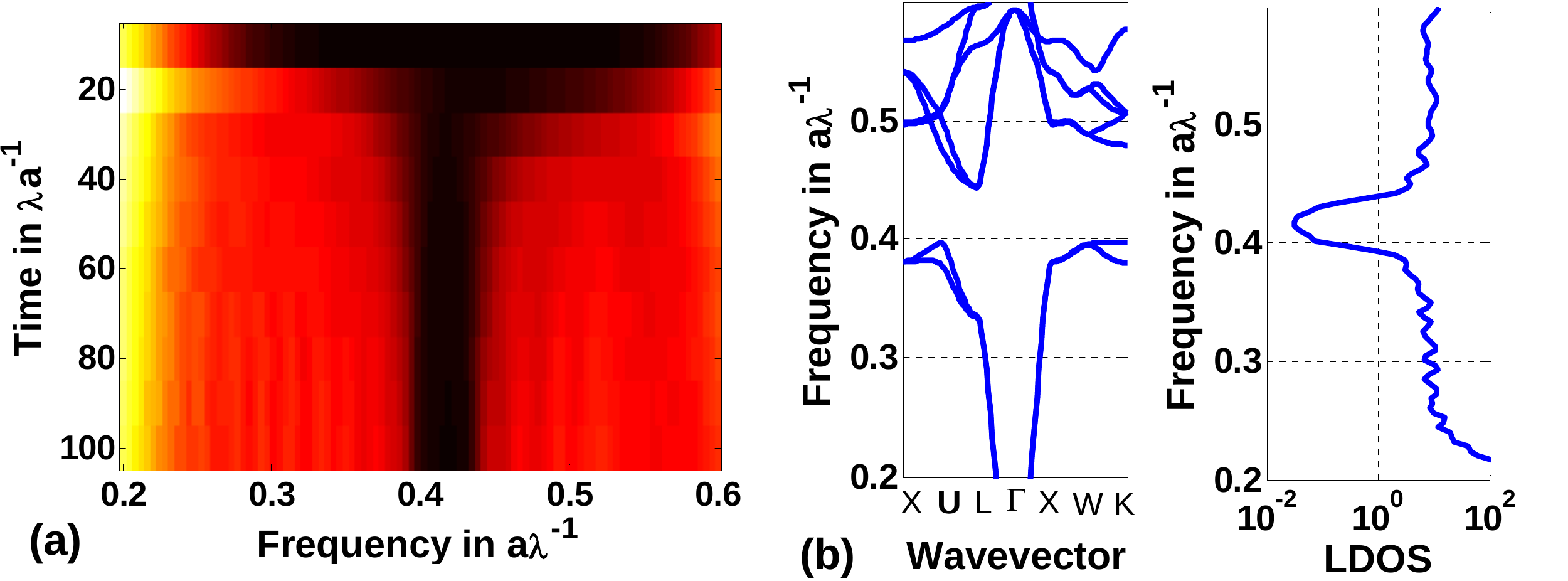}\caption[Submanifold]{(color
online) (a) Frequency dependent LDOS on a logarithmic scale for a finite
system of dielectric spheres arranged on a diamond lattice as a function of
the computational time. (b) Comparison of the band structure for the
respective infinite periodic lattice with the LDOS of the system. The
significant reduction of the LDOS clearly corresponds to the spectral region
of the complete band gap.}%
\label{Diamond_Lattice}%
\end{figure}

In a first investigation we applied our procedure to a system in
where the spheres are arranged on a diamond lattice \cite{Diamond}.
The purpose is to verify our computational approach as well as to
investigate the appearance of the LDOS in the presence of a complete
photonic band gap. The results are compared with photonic band
structure computations of the infinite lattice, proving the presence
of a complete band gap. We have chosen the spheres to have a
permittivity of $\varepsilon$~=~11.56 and a radius of $R~=~0.25$a,
with $a$ being the lattice constant. The finite diamond lattice was
defined by a $7\mathrm{a}~\times~7\mathrm{a}~\times~7\mathrm{a}$
cube. The spatial permittivity distribution was discretized in the
FDTD method with a resolution of $0.025$a. To compute the LDOS, we
located a point source at the center of the structure and computed
the time dependent transmitted power through a sphere that includes
the entire structure. The point source was excited by a temporal
Gaussian pulse. Fourier transformation of the time dependent signal
yields the frequency dependent transmission. Upon a normalization
with the power spectrum of the source, we obtain the transmission.
In a lossless medium this transmission corresponds to the LDOS
associated with the spatial point where the source was placed. On a
logarithmic scale the spectrum of the LDOS as a function the
normalized computational time is shown in Fig.~\ref{Diamond_Lattice}
(a).

After a sufficient computational time of $\approx~100~\lambda
\mathrm{a}^{-1}$ all transient effects are faded away. The LDOS
converges to a stationary distribution. The logarithmic scale
clearly reveals a strong suppression of the LDOS in a frequency
interval between 0.4 and 0.44a$\lambda^{-1}$. The main features of
the LDOS are independent on the exact position of the point source
within a unit cell. Comparing the LDOS with the band structure
(Fig.~\ref{Diamond_Lattice} (b)) clearly reveals that the spectral
region of a suppressed LDOS coincides with the band gap. For a
spatial extension of the structure of $7\mathrm{a}$ in all
dimensions, the LDOS is decreased by approximately two orders of
magnitude when compared to spectral regions for which propagation of
light is allowed.

\begin{figure}[h]
\centering
\includegraphics[width=180mm,angle=0] {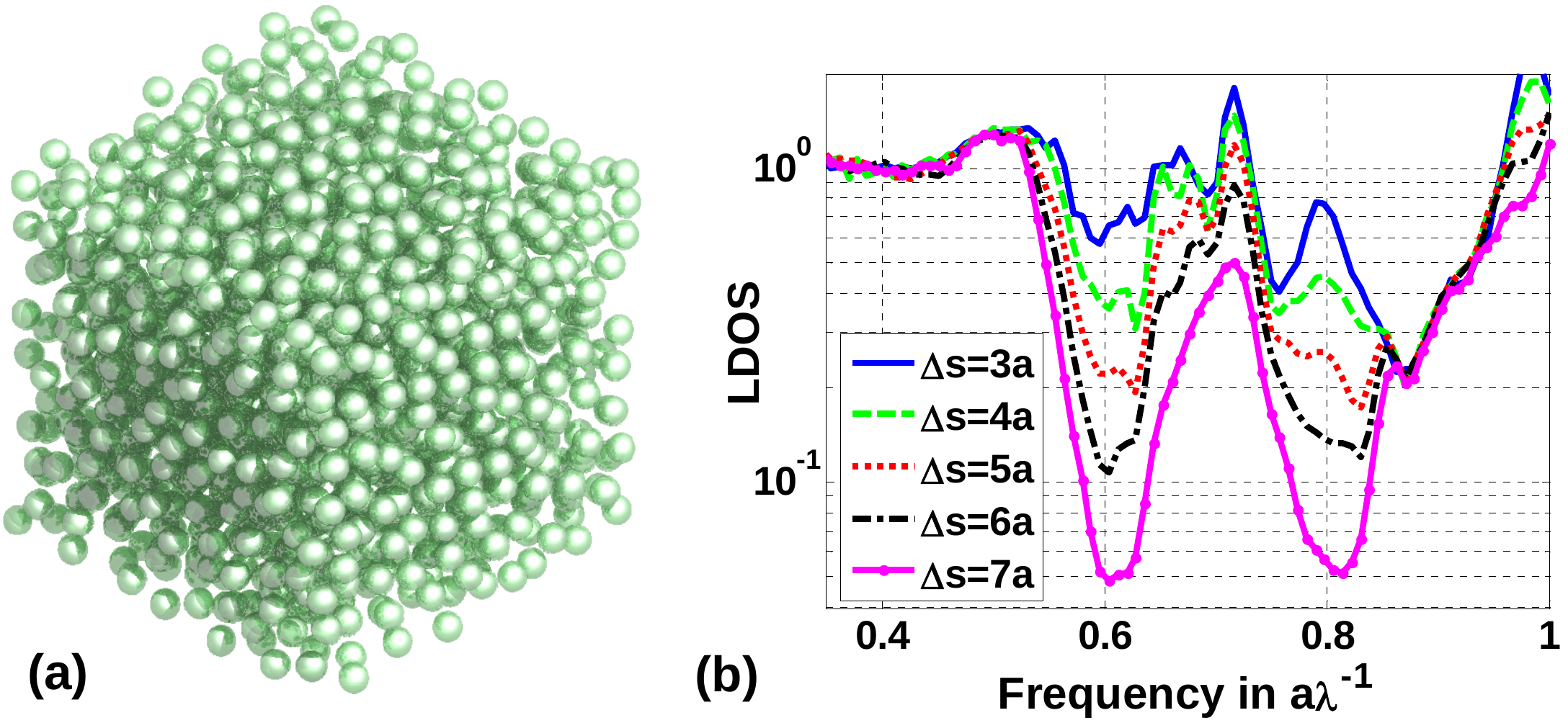}\caption[Submanifold]{(color
online) (a) Example of a random arrangement of spheres for which the LDOS was
computed. (b) LDOS as a function of the spatial extension of the domain where
spheres were randomly arranged.}%
\label{Random_size}%
\end{figure}

After having identified how band gaps and reduced LDOS are related,
we proceed in analyzing random arrangements of the same spheres
where we keep the filling fraction of the spheres constant at
$41\%$. Although a ceases to have a meaning of being the period we
use it further as the length scale. Exemplarily,
Fig.~\ref{Random_size} (a) shows a particular arrangement of the
spheres. The arrangement was achieved by generating at first
randomly a coordinate for a new sphere. Subsequently it was verified
if a sphere is already present in the volume to be occupied by the
new sphere. If this is excluded, the position is accepted, otherwise
it is rejected. The procedure is repeated until the predefined
particle density is met. The exclusion of penetrating spheres was
necessary to avoid that resonances associated with the spherical
shape of the particles (Mie resonances) were appreciably altered.
Moreover, as a result of preliminary numerical experiments, nearly
touching spheres had to be excluded. This was ensured by defining
the forbidden volume slightly larger. Nearly touching spheres are
sufficiently strong coupled to lift the degeneracy of the spectral
position of a Mie resonance, causing an inhomogeneous line
broadening and a damping of the resonance of the ensemble. As a
consequence, the LDOS would not have been sufficiently suppressed.
Hence placing a new sphere is inhibited within a radius of $0.3$a of
an existing sphere. This seemingly limitation will be subsequently
lifted by covering the spheres in the simulation with a core
material having a low dielectric constant. Such arrangement permits
for a dense packing of the spheres but maintaining the
characteristics of the LDOS.

The LDOS of a medium composed of randomly arranged spheres as a function of
the spatial extension of the domain that hosts the spheres is shown in
Fig.~\ref{Random_size} (b). At low frequencies the LDOS remains constant,
indicating that light flows off the source. The source is always located in
the center of the spatial domain of interest. On the contrary, at slightly
elevated frequencies two bands exist were the LDOS shows the exponentially
decaying strength with increasing linear size of the system. It suggests that
the transport velocity of light is strongly reduced. The two bands are
centered at $\approx$~0.59 a$\lambda^{-1}$ and at $\approx$~0.81 a$\lambda^{-1}%
$, respectively. By using different random arrangements slight
modifications in the LDOS were encountered, though the principal
features of a suppressed LDOS in the same spectral domains was
always observed.

\begin{figure}[h]
\centering
\includegraphics[width=180mm,angle=0] {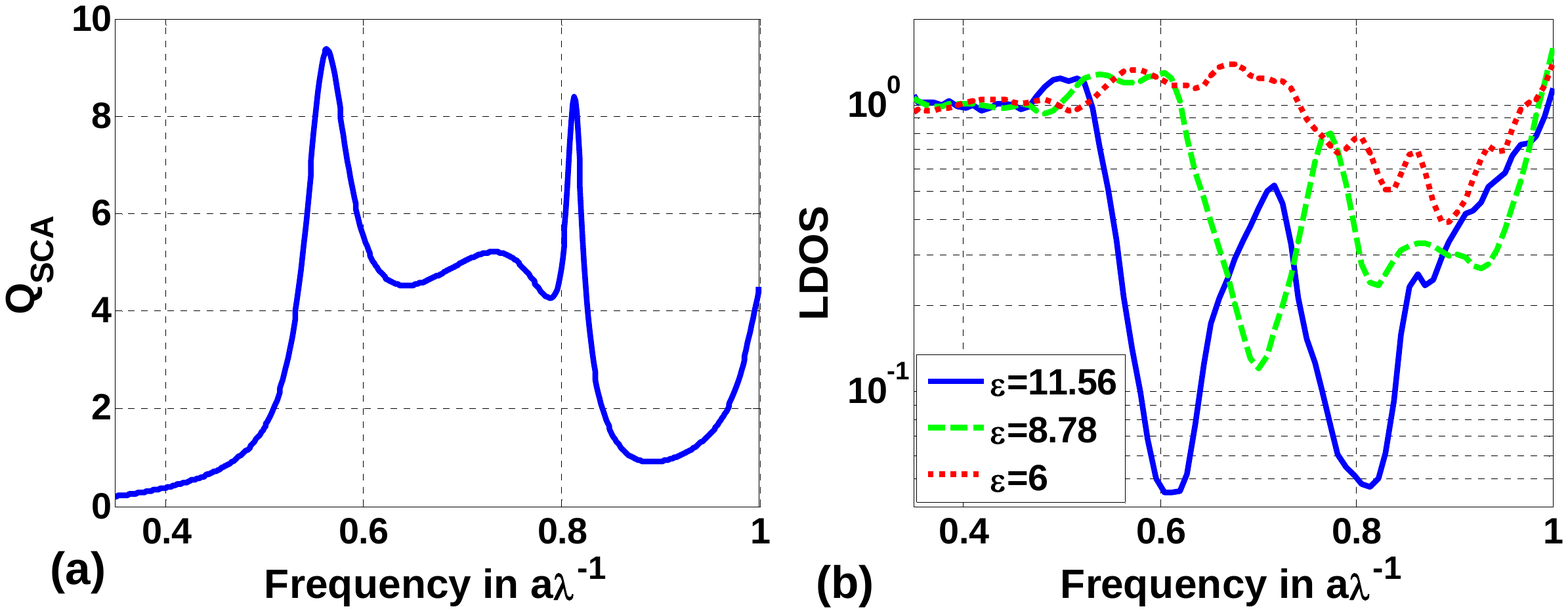}\caption[Submanifold]{(color
online) (a) Scattering cross section calculated using Mie theory for
a single sphere of radius $R~=~0.25$a and permittivity
$\varepsilon=11.56.$ The sphere is embedded in air and illuminated
by a plane wave. (b) LDOS for randomly
arranged spheres as a function of the sphere's permittivity.}%
\label{Random_epsilon_Qsca}%
\end{figure}

Both peculiar frequencies can be correlated to Mie resonances of the
isolated dielectric sphere \cite{Bohren}. For this purpose the
scattering cross section is shown in Fig.~\ref{Random_epsilon_Qsca}
(a). Two sharp resonances can be seen at
$0.56\mathrm{a}\lambda^{-1}$ and at $0.81\mathrm{a}\lambda^{-1}$.
From the Mie coefficients it can be deduced that they correspond to
resonances of the lowest eigenmodes of the electric $\left(
c_{0}\right)$ and the magnetic $\left( d_{0}\right)$ potential
inside the sphere, respectively. The reason why the suppression of
light transport occurs at slightly larger frequencies compared to
the Mie resonances lies potentially in the requirement that the
field scattered by the sphere oscillates $\pi$ out of phase with
respect to the incident field. This will effectively suppress the
propagation of light in the forward direction since scattered and
illuminating fields annihilate due to destructive interference
provided that the scattering strength of either sphere is
sufficiently strong. This required phase difference is encountered,
as in all resonance phenomena, slightly above the resonance
frequencies.

Modifying the permittivity of the sphere will have various
implications. Foremost, the spectral position of the resonances
shifts towards larger frequencies, the resonance gets weaker and
broader, and overall, the scattering strength is reduced. These
modifications of the LDOS for randomly arranged spheres are shown in
Fig.~\ref{Random_epsilon_Qsca} (b). The spatial extension was chosen
to be $(7\mathrm{a})^{3}$. The spectral regions of the reduced LDOS
shifts to higher frequencies being in perfect agreement with
prediction of Mie's theory. Furthermore, the anticipated smaller
reduction of the LDOS for a smaller permittivity is evident. If the
permittivity of the spheres is smaller than $\varepsilon\approx6$,
defining a domain of suppressed LDOS is pointless.

Because the structures investigated up to now cannot be fabricated,
we now investigate core-shell particles. The shell is made of a
material with a sufficiently low permittivity. It should not affect
the strength of the scattering resonance of the core. Furthermore,
the shell should have an appropriate thickness to allow for a
sufficient separation of the core spheres. As mentioned we have
chosen a core-shell radius of $0.3$a with a shell permittivity of
$\varepsilon~=~2.25$. The core features were kept invariant
($R~=~0.25$a, $\varepsilon~=~11.56$). The self sustaining ensemble,
reasonable for fabrication, was generated with the sphere dropping
method. In this method a sphere is dropped into a fictitious box.
Gravity forces the sphere to sink down (along the $z$-axis) to the
bottom of the box. If the dropped sphere touches a another sphere a
translation in the $x-y$-plane of the dropped sphere is performed.
The translation is chosen such that the distance between the dropped
sphere and the other spheres is maximized. It allows to sink the
dropped sphere further down until it touches again another sphere.
The procedure is repeated until the dropped sphere cannot sink
further, although it is incrementally replaced in the $x-y$-plane in
an arbitrary direction. Once this stage in the procedure is reached,
the sphere is stably trapped by its surrounding.

\begin{figure}[h]
\centering
\includegraphics[width=180mm,angle=0] {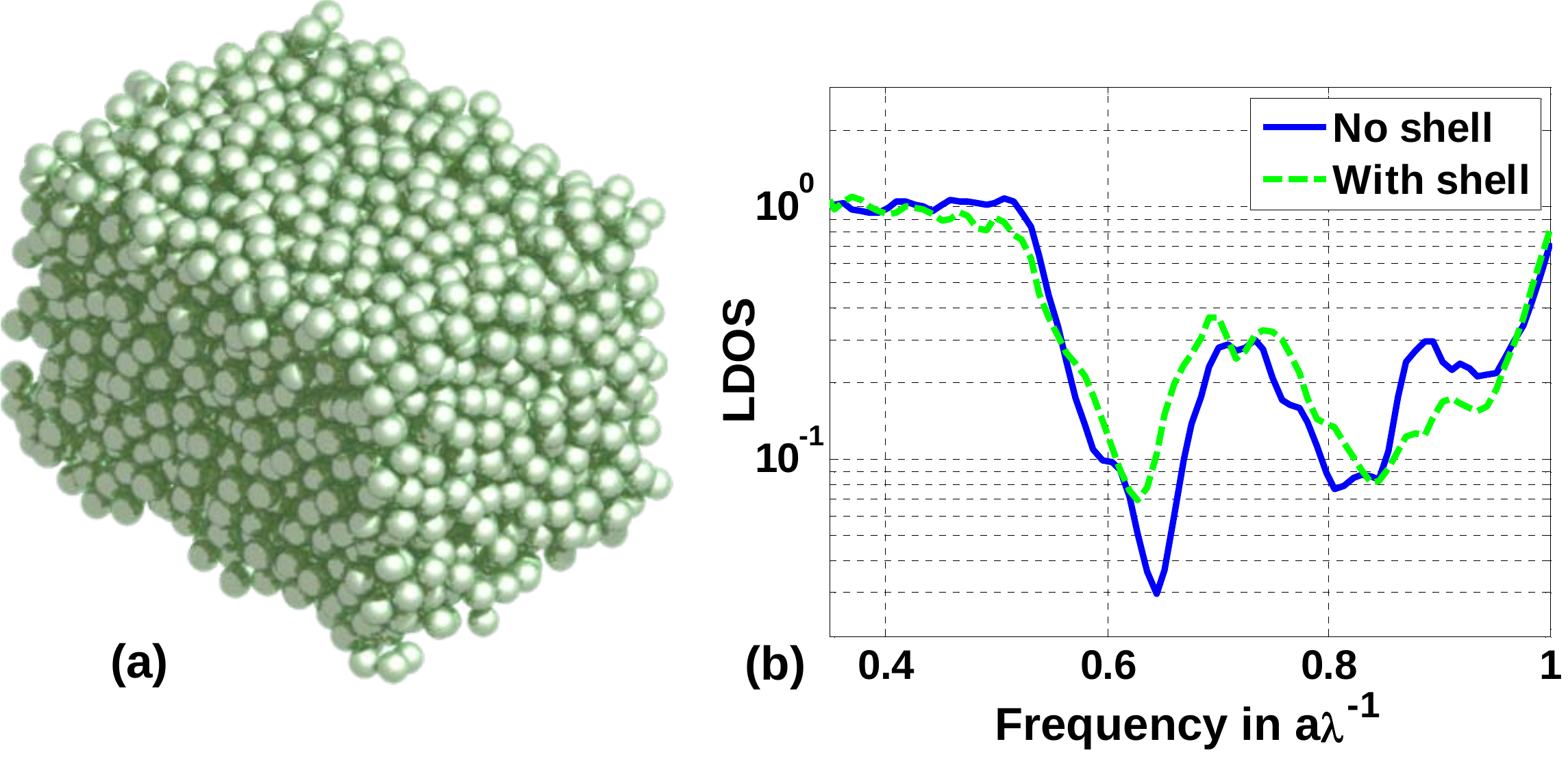}\caption[Submanifold]{(color
online) (a) Spatial arrangement of a self sustaining ensemble of randomly
arranged spheres as generated by the sphere dropping method. (b) LDOS for
cores with or without shells. The arrangement is only self sustaining if the
spheres consist of cores and shells.}%
\label{Core_Shell}%
\end{figure}

The procedure allows to generate self-sustaining ensembles of
randomly arranged spheres with structural integrity. To force the
randomness we assumed hard boundaries. The final spatial extension
of the structure is the same as before and amounts to
$(7\mathrm{a})^{3}$. On passing we note that in small spatial
domains a certain periodicity is encountered as the spheres tend to
form an opal structure. But as no energy minimum for the global
arrangement was forced, the structure will retain its random
character in general. A distribution of spheres generated with this
method is shown in Fig.~\ref{Core_Shell} (a). The filling fraction
amounts to 54\%.

The corresponding LDOS for such a core-shell structure is shown in
Fig.~\ref{Core_Shell} (b). For the sake of comparison we show also
the LDOS for the structure without shells. In this case the larger
density of spheres results in a stronger suppression of the LDOS
when compared to the scenario analyzed before. However, the spectral
position of the suppression dips coincide. The optically weak shell
causes primarily a slightly increased LDOS in the spectral domain of
interest, though it remains some orders of magnitude smaller
compared to spectral domains where light propagation is allowed. A
slight shift of the spectral domains of reduced LDOS appears for the
second resonance frequency of $0.805\mathrm{a}\lambda^{-1}$. Thus we
conclude that the shell does not significantly affect the properties
of the photonic material, though a minor degradation can be
observed. On the other hand, the shell is sufficiently thick to
guarantee the structural integrity and to make the fabrication of
the photonic material feasible.

Such compactness is extremely important if this photonic material
shall be used in applications. The omnidrectional gap can be
exploited, e.g., by using this material as intermediate reflector in
tandem solar cells. The problem to be solved in such tandem solar
cells is to define an intermediate reflector that reflects all light
in a certain frequency interval and independent of the angle of
incidence back into the top cell, leaving the other spectral
components unaffected. Maximizing the light absorbance in the top
cell allows to increase the overall efficiency of the solar cell.
The proposed photonic material is very promising for such an
application, as it acts sufficiently efficient and can be fabricated
at low costs.

In conclusion, we have investigated the local density of states for
a photonic material made of randomly arranged spheres. It was shown
that a significant suppression of the LDOS is encountered in
frequency domains slightly above the Mie resonances of the single
sphere. The observation of the effect requires a minimum distance to
adjacent spheres. The magnitude of the LDOS decays exponentially
with a linearly increasing size of the photonic material indicating
the presence of a photonic band gap of the infinite structure.
Covering the spheres with a shell of a permittivity as small as
feasible allows to observe the same phenomena for self-sustaining
ensembles. The theoretical concept might be a way to fabricate  a
photonic material at low costs that suppresses the light propagation
along all directions, representing  an omnidirectional band gap
material.

We would like to acknowledge the partial financial support of this work
through the Deutsche Forschungsgemeinschaft (PAK88) and the Federal Ministry
of Education and Research (Nanovolt). Some computations utilized the IBM p690
cluster JUMP of the Forschungszentrum in Jülich, Germany.

\end{document}